\documentclass[10pt]{article}
\usepackage{fullpage}
\usepackage{amsfonts}
\usepackage{amsmath}
\usepackage{amsthm}
\usepackage{graphicx}
\usepackage{color}
\usepackage{amssymb}
\usepackage{empheq}
\usepackage{mathrsfs}
\usepackage{enumerate}
\usepackage{tikz}
\usepackage{pgflibraryarrows}
\usepackage{pgflibrarysnakes}
\usepackage{upgreek}
\usepackage{tipa}
\usepackage{multicol}
\usepackage{verbatim}
\usepackage{caption}

\newcommand{\de}[2]{\frac{d #1}{d #2}}
\newcommand{\pd}[2]{\frac{\partial #1}{\partial #2}}

\title{Host-feeding enhances stability of discrete-time host-parasitoid population dynamic models}
\author{Brooks Emerick\footnote{Department of Mathematics, Trinity College, Hartford CT, 06106}, Abhyudai Singh\footnote{Department of Electrical and Computer Engineering, University of Delaware, Newark DE, 19716}$\,\,^,$\footnote{Department of Biomedical Engineering, University of Delaware, Newark DE, 19716}}

\begin{document}
\maketitle
\author



%
%
%

\begin{abstract}
Discrete-time models are the traditional approach for capturing population dynamics of a host-parasitoid system. Recent work has introduced a semi-discrete framework for obtaining model update functions that connect host-parasitoid population levels from year-to-year. In particular, this framework uses differential equations to describe the hosts-parasitoid interaction during the time of year where they come in contact, allowing specific behaviors to be mechanistically incorporated into the model. We use the semi-discrete approach to study the effects of host-feeding, which occurs when a parasitoid consumes a potential host larva without ovipositing. Our results show that host-feeding by itself cannot stabilize the system, and both the host and parasitoid populations exhibit diverging oscillations similar to the  Nicholson-Bailey model.  However, when combined with other stabilizing mechanisms such as density-dependent host mortality or density-dependent parasitoid attack rate, host-feeding expands the region of parameter space that allows for a stable host-parasitoid equilibrium. Finally, our results show that host-feeding causes inefficiency in the parasitoid population, which yields a higher population of hosts per generation. This suggests that host-feeding may have limited long-term impact in terms of suppressing host levels for biological control applications.   

\end{abstract}







\section{Introduction}
The host-parasitoid dynamic typically involves a vulnerable period during the year in which hosts are susceptible to attack by parasitoids.  There exists a tendency of synovigenic parasitoids to eat the host without laying eggs inside \cite{Hawkins_1994, Hawkins_1994b}.  Generally, an adult female parasitoid emerges every year with less eggs than she can potentially oviposit in her lifespan.  Therefore, a parasitoid will feed on hosts to gain the necessary energy and tissue to mature additional eggs \cite{Jervis_1986, Kidd_1989}.  However, in doing so, the parasitoid loses a potential host as the majority of parasitoids kill the host during host-feeding.  This results in a high death rate of the host population during the vulnerable period, which can have interesting consequences on later generations \cite{Bach_1943, Flanders_1953}.  For further reviews on the host-feeding interaction and its biological implications, we refer the interested reader to \cite{Jervis_1986, Jervis_1996, Ueno_1998}.  

A traditional approach to describe parasitoid-host dynamics is to use discrete time models, such as the Nicholson-Bailey model \cite{Nicholson_1935} and others considered more recently \cite{Bompard_2013, Jang_2012, Jones_2011, Cobbold_2009, Kapcak_2013}.  Discrete models can monitor the change in population density for distinct points in time, such as each year or each generation, which agrees with typical life-cycle themes of insects in temperate climates \cite{Murdoch_2003}.  Several authors incorporate the effects of host-feeding in continuous models, which are suitable to year-round interactions such as those in tropical climates.  They conclude that host-feeding can have stabilizing effects \cite{Yamamura_1988, Murdoch_1992} or no effect on stability \cite{Briggs_1995, Kidd_1991, Kidd_1991b}, while Shea et.~al.~conclude that egg production delay has a destabilizing effect in host-feeding models \cite{Shea_1996}.  Further reviews on past and recent models are provided by Hassell et.~al.~in \cite{Hassell_2000b} and \cite{Hassell_2000} and by Murdoch et.~al.~in \cite{Murdoch_2003}.

In this work, we use the semi-discrete framework \cite{Pachepsky_2007} to develop discrete update functions based on a continuous host-feeding dynamic.  This modeling approach is more mechanistic as compared to early phenomenological models because it incorporates specific host-parasitoid interactions during the vulnerable period.  We conclude that host-feeding has no effect on stability as compared to previous results by Singh et.~al~\cite{Singh_2007}; however, host-feeding combined with density-dependent mortality rates and functional responses has a stabilizing effect.  The paper is organized as follows: we formulate the model in Section \ref{sec2}; in Section \ref{sec3}, we consider the simplest case and show that host-feeding alone does not stabilize the Nicholson-Bailey model \cite{Nicholson_1935}; in Sections \ref{sec4} and \ref{sec5}, we incorporate the effects of density-dependent mortality and quadratic functional response, respectively, and consider the stability of the system; we conclude with several discussion topics in Section \ref{sec6}.  




\section{Model Formulation}\label{sec2}
\begin{figure}[h!]
\centering
\includegraphics[width = .65\textwidth]{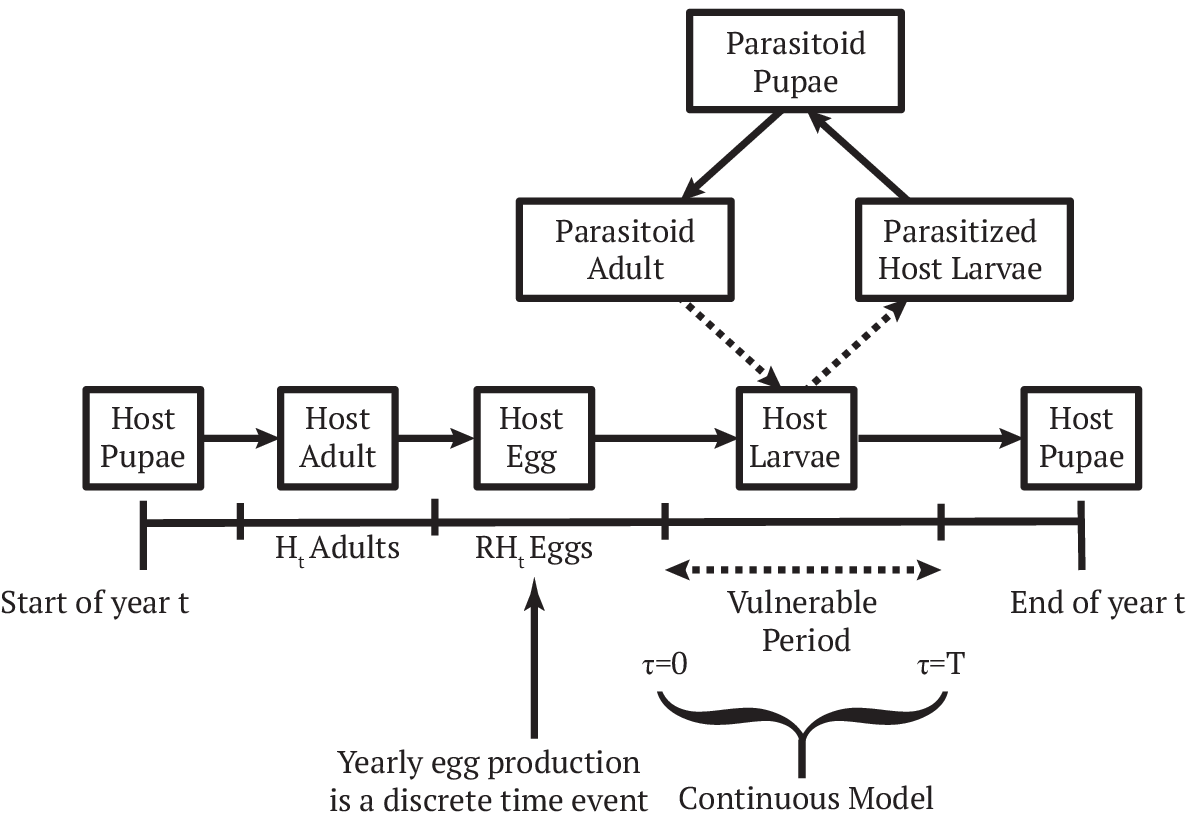}
\captionof{figure}{Life cycle of the host and parasitoids in year $t$.}
\label{life_cycle}
\end{figure}

Generally, the discrete-time model used to describe the host-parasitoid dynamics is given by 
\begin{align}
H_{t+1} & = F(H_t,P_t) \label{H_plus_1}\\
P_{t+1} & = G(H_t,P_t) , \label{P_plus_1} \end{align}
where $H_t$ and $P_t$ are the adult female host and parasitoid densities, respectively, at the beginning of the vulnerable period of year $t$, where $t$ is an integer.  Figure \ref{life_cycle} illustrates the life cycle of the host and parasitoids in a year long period.  During year $t$, host pupae mature into $H_t$ adults.  These $H_t$ adults lay a particular amount of eggs, which eventually mature into $RH_t$ larvae at the beginning of the vulnerable stage.  Here, $R>1$ denotes the number of viable eggs produced by each adult host.  The time within the vulnerable stage is denoted by $\tau$, where $\tau\in[0,T]$.  Time $\tau = 0$ denotes the beginning of the vulnerable stage and $\tau = T$ denotes the end.  During this time, adult female parasitoids emerge with egg and seek out host larvae to oviposit inside them.  The parasitoid egg inside the host becomes a juvenile parasitoid that uses the host as a resource.  At the end of the vulnerable period, a proportion of the host population is infected.  This results in an initial parasitoid population the next year as juvenile parasitoids emerge from the host to continue their life cycle.  A number of hosts escape parasitism, and survive to mature into pupae, which pupate over the winter creating the next year's population of adult hosts.  To better understand the discrete update of each year, we consider a continuous model to describe the interaction of hosts and parasitoids during the vulnerable period.  Next, we discuss the approach to model this system of interactions.  

\subsection{The Semi-Discrete Framework}\label{sec2_1}
A continuous time model is used to describe the dynamics of the interacting host and parasitoid populations during the vulnerable period.  The update functions $F$ and $G$ of the discrete model depend on the output of the continuous model at the end of the vulnerable period each year.  We consider the following chemical reaction 
\begin{align}
&P + L \xrightarrow{g(\,\cdot \,)} I +P. \end{align}

\noindent Here, $g(\,\cdot \,)$ (units: $time^{-1}\, parasitoid^{-1}$) is the attack rate of the parasitoids, which represents the instantaneous rate at which the hosts are attacked per parasitoid.  This function could potentially be dependent on the population of hosts, parasitoids, or infected hosts.  In general, we write the continuous model as
\begin{align}
\de{L(\tau,t)}{\tau} & = -g(\,\cdot\,)L(\tau,t)P(\tau,t) \\
\de{I(\tau,t)}{\tau} & = g(\,\cdot\,)L(\tau,t)P(\tau,t) \\
\de{P(\tau,t)}{\tau} & = 0.\end{align}
where $L(\tau,t)$, $I(\tau,t)$, and $P(\tau,t)$ denote the concentrations of host larvae, infected larvae, and parasitoids, respectively, at time $\tau$ during the vulnerable period in year $t$.  We subject the system to the following initial conditions 
\begin{align}
& L(0,t) = RH_t, \quad I(0,t) = 0, \quad P(0,t) = P_t. \end{align}
These equations are integrated from $\tau = 0$ to $\tau = T$.  Assuming each parasitized host larvae gives rise to $k$ adult parasitoids in the next generation, the update functions in the discrete model are 
\begin{align}
H_{t+1}=F(H_t, P_t) & : = L(T,t)  \label{Update_F} \\
P_{t+1}=G(H_t, P_t) & : = kI(T,t). \label{Update_G} \end{align}
\noindent Once the system is formulated, investigation into the stability region of Equations \eqref{H_plus_1} and \eqref{P_plus_1} can be carried out using the standard Jury conditions \cite{Elaydi_1996}.  If we let $g = c$, i.e. a constant parasitoid attack rate, then we obtain the following update
\begin{align}
F(H_t, P_t) & = RH_t \exp(-cP_tT)\\
G(H_t, P_t) & = kRH_t\big[1-\exp(-cP_tT)\big], \end{align}
which is the classic Nicholson-Bailey model.  This interaction is unstable and over time both populations experience diverging oscillations \cite{Murdoch_2003}.  Several changes in the form of functional responses, susceptibility of attack, and density-dependent mortality can be made to stabilize the system as in \cite{Singh_2007, Singh_2009}.  Next, we illustrate how we incorporate host-feeding into the semi-discrete framework and discuss how to obtain the update functions.  


\subsection{The Semi-Discrete Framework with Host-Feeding}\label{sec2_2}

The continuous model we consider includes the tendency of the parasitoids to feed on host larvae without infecting them with an egg \cite{Jervis_1986, Kidd_1989}.  This means the parasitoid population has two phases: without egg and with egg.  If a parasitoid lays an egg inside the larva, an infected larva is produced and the parasitoid is now without an egg.  This eggless parasitoid must feed on a host larva before gaining another egg; however, if a parasitoid has an egg, it will infect the larvae rather than feed.  Implicit to our model is the assumption that once a parasitoid feeds on the host, it immediately gains enough energy to produce an egg, i.e., the transitional period from eggless to with egg is instantaneous; however, we do discuss consequences of a delay in Section \ref{sec6}.  This approach is motivated by the work of Shea et.~al. \cite{Shea_1996}, who consider several parasitoid phases with an egg maturation delay.  A kinetic diagram below describes the process, 
\begin{align}
&P_1 + L \xrightarrow{g(\, \cdot \,)} I + P_0\label{Kinetic_1}\\
& P_0 + L \xrightarrow{g(\, \cdot \,)} P_{1}. \label{Kinetic_2}\end{align}
Here, the function $g(\, \cdot \,)$ (units: $time^{-1}\, parasitoid^{-1}$) represents two means by which the parasitoids attack hosts.  The first reaction is the instantaneous rate at which the hosts are parasitized or infected by a mature parasitoid.  The second reaction represents the rate at which the hosts are devoured by an eggless parasitoid.  We assume the two rates are equal as this leads to analytic results.  In the most general case, both rates may potentially depend on the host, parasitoid, or infected host populations, which yields the following continuous model
\begin{align}
\de{L(\tau,t)}{\tau}  &= -g(\,\cdot\,) \big(P_0(\tau,t) + P_1(\tau,t) \big) L(\tau,t) \label{ODE_1} \\
\de{I(\tau,t)}{\tau} & = g(\,\cdot\,)P_1(\tau,t) L(\tau,t)\label{ODE_2} \\
\de{P_0(\tau,t)}{\tau}  & = g(\,\cdot\,)\big(P_1(\tau,t) - P_0(\tau,t)\big) L(\tau,t)\label{ODE_3} \\
\de{P_1(\tau,t)}{\tau}  & =  -g(\,\cdot\,)\big(P_1(\tau,t) - P_{0}(\tau,t)\big)L(\tau,t),\label{ODE_4}\end{align}

\noindent where $L(\tau,t)$, $I(\tau,t)$, $P_0(\tau,t)$, and $P_1(\tau,t)$ are the density of host larvae; parasitized host larvae; eggless parasitoids;  and parasitoids with egg, respectively, at time $\tau$ during the vulnerable period and in year $t$.  We subject the system to the following initial conditions 
\begin{align}
& L(0,t) = RH_t, \quad I(0,t) = 0, \quad P_0(0,t) = 0,\quad P_1(0,t) = P_t. \end{align}
This initial condition assumes all parasitoids emerge with an egg; however, we also consider a proportion of the population that emerges eggless.  We discuss the results of this assumption briefly in the conclusion.  These equations are integrated from $\tau = 0$ to $\tau = T$.  Assuming each parasitized host larvae gives rise to $k$ adult parasitoids in the next generation, the yearly update functions are the same as before in Equations \eqref{Update_F} and \eqref{Update_G}.  We note here that $P_t = P_0(0,t) + P_1(0,t)$ if there is no density-dependent parasitoid mortality.  For the remainder of the paper, we suppress the dependence on $\tau$ and $t$ for the functions $L$, $I$, $P_0$, and $P_1$ for convenience.  

\section{Nicholson-Bailey with Host-Feeding}\label{sec3}
In this section, we consider the most simple case.  We assume the attack rate $g(\,\cdot\,)$ in Equations \eqref{Kinetic_1} and \eqref{Kinetic_2} is constant, i.e. $g = c$.  Our continuous time model during the vulnerable period from Equations \eqref{ODE_1} -- \eqref{ODE_4} becomes
\begin{align}
\de{L}{\tau} & = -c(P_0+P_1)L \label{No_Mort_1}\\
\de{I}{\tau} & = cP_1 L   \label{No_Mort_2}\\
\de{P_0}{\tau} & = c(P_1-P_0)L   \label{No_Mort_3}\\
\de{P_1}{\tau} & = -c(P_1 - P_0) L. \label{No_Mort_4}\end{align}
\noindent In this case, we can solve the system for $L$ and $I$ explicitly (\ref{appA}) and our discrete update from Equations \eqref{H_plus_1} and \eqref{P_plus_1} can be written as
\begin{align}
H_{t+1} & = RH_t e^{-cP_t T} \label{No_Mort_Update_H}\\
P_{t+1} & = \frac{kRH_t}{2} \big( 1 - e^{-cP_t T} \big) +  \frac{k P_t}{2}\, \left\{ \text{exp}\left[ \frac{RH_t}{P_t} \big( 1 - e^{-cP_tT} \big) \right] - 1 \right\}  \label{No_Mort_Update_P}. \end{align}
Here, without loss of generality and for the remainder of the paper, we assume $T = 1$ and $k = 1$ since these parameters have only a scaling effect on the results.  Here, $T = 1$ corresponds to one vulnerable period, which is approximately 90 days.  The fixed points to this model satisfy
\begin{equation}H^* = \frac{\beta \ln R}{c(R-1)}, \qquad P^* = \frac{\ln R}{c}, \qquad \beta = 0.7921.\end{equation}
\noindent As shown in \ref{appA}, this model is unstable for all $R$.  Even if we introduce a transition rate from $P_0$ to $P_1$ via an egg maturation delay, the model is still unstable.  This suggests that host-feeding alone cannot establish stability, which is in agreement to previous models \cite{Briggs_1995, Kidd_1991}. 

\section{Density-dependent Host Mortality with Host-Feeding}\label{sec4}

\begin{figure}[t!]
\centering
\includegraphics[width=\linewidth]{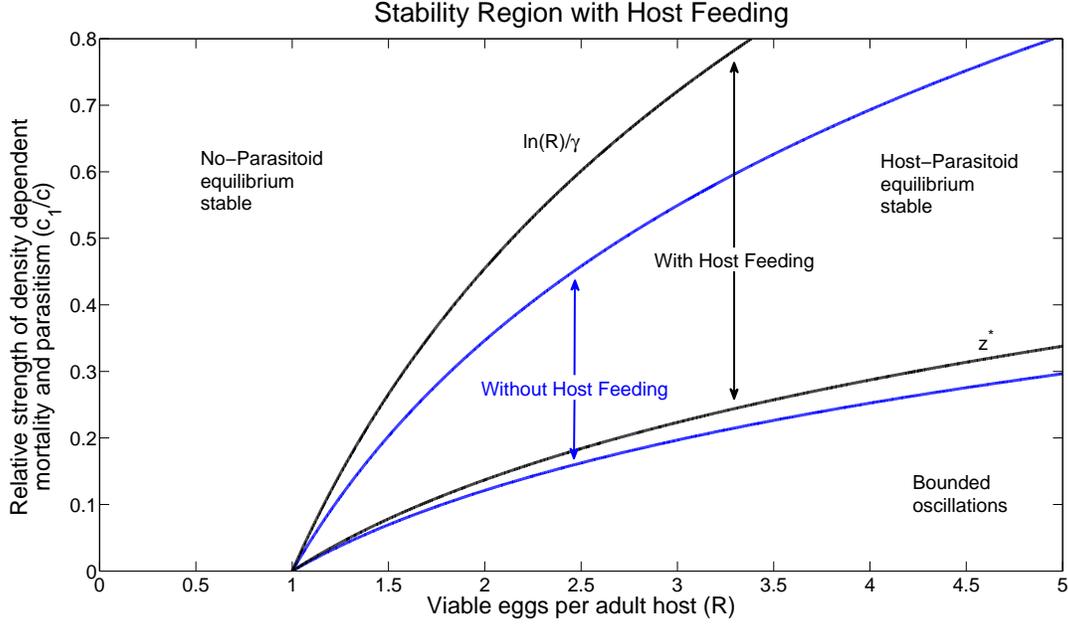}
\captionof{figure}{\textbf{Host-feeding increases the stability region of the Nicholson-Bailey model with density-dependent mortality}.  The stability region specified in \eqref{Z_star_ineq}, for the discrete time, host mortality model (Equations \eqref{Mort_Update_H} and \eqref{Mort_Update_P}) as a function of the strength of density-dependent mortality to parasitism $(c_1/c)$ and the number of viable eggs per host, $R$ (black line).  The stability region is larger with host-feeding, as compared to the density-dependent host mortality model without host-feeding (blue line) \cite{Singh_2007}. }
\label{lambda005_plot}
\end{figure}

It is shown in both phenomenological models \cite{May_1981, Hassell_1969} and mechanistic approaches \cite{Singh_2007} that density-dependent host mortality can stabilize discrete-time systems.  To investigate stability with host-feeding, we implement a density-dependent host mortality in Equations \eqref{ODE_1} -- \eqref{ODE_4}.  The kinetic reactions of the system are depicted as 
\begin{align}
&P_1 + L \xrightarrow{g(\, \cdot \,)} I + P_0\\
& P_0 + L \xrightarrow{g(\, \cdot \,)} P_{1} \\
& L \xrightarrow{g_1(\, \cdot \,)}\text{Death}.\end{align}
The function $g_1(\, \cdot \,)$ (units: $time^{-1}$) represents the (potentially density-dependent) host mortality rate due to causes other than parasitism and host-feeding.  We assume the attack rates are constant, $g = c$, implying a linear functional response.  We let the host mortality rate depend on the current amount of host larvae, i.e., $g_1 = c_1 L$ so that density-dependent effects can act simultaneously with parasitism in the continuous-time model.  We obtain the following system for $\tau\in [0,T]$, 
\begin{align}
\de{L}{\tau} & = -c(P_1 + P_0)L - c_1 L^2  \label{Mort_1}\\
\de{I}{\tau} & = cP_1L \label{Mort_2}\\
\de{P_0}{\tau} & = c(P_1 - P_0)L  \label{Mort_3}\\
\de{P_1}{\tau} & = -c(P_1 - P_0)L, \label{Mort_4}\end{align}
subject to the same initial conditions as above.  Solving this system explicitly (\ref{appB}) yields the following discrete update system 
\begin{align}
H_{t+1}  =&\,\, \frac{RH_t \exp(-cP_t)}{f(H_t, P_t) } \label{Mort_Update_H}\\
P_{t+1}  =&\,\, \frac{cP_t}{2c_1} \ln \left[ f(H_t, P_t) \right] +\frac{P_t}{4}\left[  f(H_t, P_t)^{-\frac{2c}{c_1}}-1\right],\label{Mort_Update_P}\end{align}
where 
\begin{equation} f(H_t,P_t) =  1 + c_1RH_t \frac{1-\exp(-cP_t )}{cP_t}. \end{equation}
\noindent We obtain two nontrivial fixed point solutions.  The no-parasitoid (NP) fixed point equilibrium is given by 
\[ H^*_{NP} = \frac{R-1}{c_1 R}, \qquad P^* = 0.\]
The stability analysis in \ref{appB} shows that the no-parasitoid equilibrium is stable for 
\begin{equation} \frac{\ln R}{\gamma} < \frac{c_1}{c}.\end{equation}
We can see from this that sufficiently large values of $c_1/c$, which describes the strength of density-dependent mortality versus parasitism, stabilizes the no-parasitoid equilibrium.  The second equilibrium point characterizes a presence of both hosts and parasitoids, and is given by 
\begin{equation}
H^* =  \left(\frac{\exp\left( \frac{\gamma c_1}{c} \right) - 1 }{ 1 -\frac{ \exp\left( \frac{\gamma c_1}{c} \right)}{R} }\right) \frac{ c P^* }{ c_1 R}, \qquad P^* = \frac{\ln(R) - \frac{\gamma c_1}{c} }{c}, \qquad  \gamma = 1.5238. \end{equation}
We consider the region of stable solutions for various values of $c_1/c$.  An analysis in \ref{appB} shows the equilibrium point is stable for 
\begin{equation} z^* < \frac{c_1}{c} < \frac{\ln R}{\gamma},\label{Z_star_ineq}\end{equation}
where $z^*$ satisfies the following equation  
\begin{equation}\frac{\left\{ R\left[ \left(\gamma z^* + 1\right) - \ln R\right] \left( 1 - e^{-\gamma z^*}\right) + 1 - e^{\gamma z^*} \right\} \left( 1 + e^{-2\gamma}\right) }{2 z^* e^{\gamma z^*}\left( 1 - Re^{-\gamma z^*} \right) } + \frac{1}{2} \left[ \gamma + \frac{1}{2} \left( 1- e^{-2\gamma} \right) \right] e^{-\gamma z^*} = 1.\label{Z_star_eq}\end{equation}
Figure \ref{lambda005_plot} shows the stability region as compared to Singh et.~al.'s \cite{Singh_2007} model without host-feeding.  We can conclude from this analysis that coupling host-feeding with density-dependent host mortality yields a larger stability region than that of the same model without host-feeding.  Overall, this ensures that the host-feeding dynamic has a stabilizing effect.  We note that the comparison in Figure \ref{lambda005_plot} is between the host-feeding model presented here and the Singh et.~al.~\cite{Singh_2007} host-mortality model with parasitic attack rate $c/2$.  In this sense, the infection rates of the attacking parasitoids, $P_1$, of both models are comparable.  


\section{Quadratic Functional Response with Host-Feeding}\label{sec5}
\begin{figure}[h!]
\centering
\includegraphics[width = \linewidth]{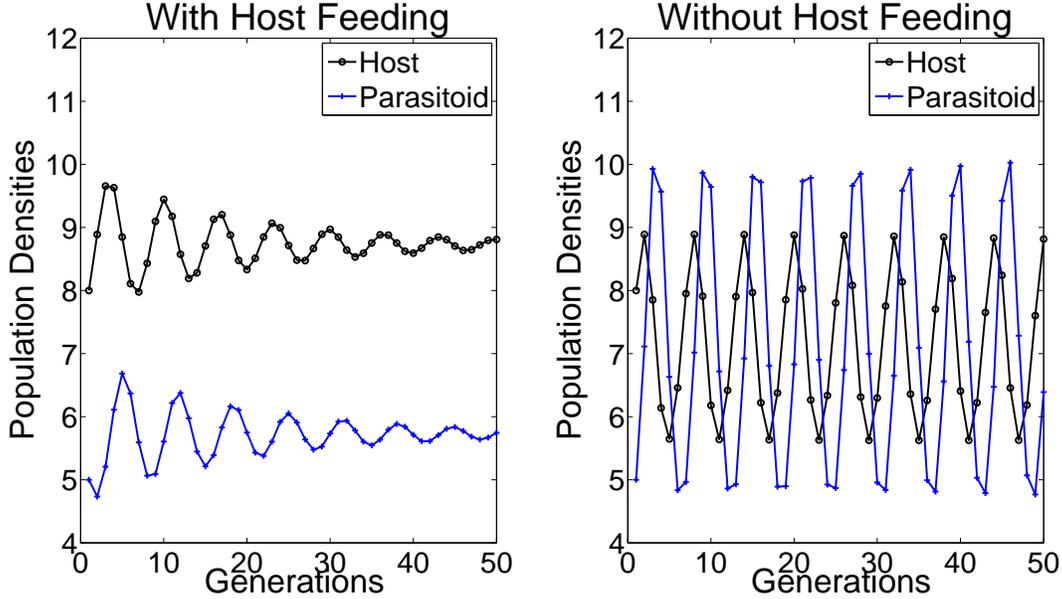}
\captionof{figure}{\textbf{Host-feeding stabilizes the Nicholson-Bailey model with quadratic functional response}.  Comparison of trajectories of the dynamic interaction given by Equations \eqref{Func_1} -- \eqref{Func_4} (left) to Singh et.~al.'s quadratic functional response model without host-feeding \cite{Singh_2007} (right).  Host-feeding with a quadratic functional response stabilizes the oscillatory behavior of Singh et.~al.'s model.  Simulations are run with $R = 2$, $c = 0.01$ and initial densities are taken as 8 and 5 for the host and parasitoid population, respectively.}
\label{Singh_Trajectory}
\end{figure}

Previous results show that phenomenological update functions with Type II and Type III functional responses do not stabilize the Nicholson-Bailey model \cite{Rogers_1972, Hassell_1978}.  However, using the semi-discrete framework, Singh et.~al.~showed that a quadratic functional response yields a neutrally stable fixed point with period $2\pi/\arctan(\sqrt{R^2-1})$ in the absence of host-feeding \cite{Singh_2007}.  To investigate the effects of host-feeding, we consider a functional response in the attack rate of Equations \eqref{Kinetic_1} and \eqref{Kinetic_2} so that $g = cL$.  Using Equations \eqref{ODE_1} -- \eqref{ODE_4}, our continuous model becomes 
\begin{align}
\de{L}{\tau} & = -c(P_0 + P_1)L^2 \label{Func_1}\\
\de{I}{\tau} & = cP_1L^2 \label{Func_2}\\
\de{P_0}{\tau} & = c(P_1-P_0)L^2 \label{Func_3}\\
\de{P_1}{\tau} & = -c(P_1 - P_0) L^2.\label{Func_4}\end{align}
subject to the same initial conditions.  In \ref{appC}, we solve this system explicitly for $L$ and $I$ to obtain the following discrete yearly update system 
\begin{align}
H_{t+1}  =&\,\, \frac{RH_t}{1 + cRH_tP_t}\label{Func_Update_H}\\
P_{t+1}  =&\,\, \frac{ RH_t - H_{t+1} + \frac{P_t}{2} \left\{ 1 - \exp\left[ -\frac{2(RH_t - H_{t+1})}{P_t} \right]\right\} }{2}.  \label{Func_Update_P} \end{align}
The fixed point of the system is
\begin{equation} H^* = \sqrt{\frac{\gamma}{cR}}, \qquad P^* = \frac{R - 1}{\sqrt{\gamma cR}}, \quad \gamma = 1.5238.\end{equation}
Stability analysis (\ref{appC}) shows that this fixed point is stable for all $R$.  Figure \ref{Singh_Trajectory} illustrates a simulation for $R = 2$ and $c = 0.01$ with and without host-feeding.  We can see that without host-feeding, the populations oscillate forever, but with host-feeding, the populations settle to a limiting value.  Hence, including host-feeding with a quadratic functional response in both the attack and infection rate stabilized the originally neutrally stable fixed point.  It should also be noted that the above host equilibrium is higher than the corresponding quadratic functional response equilibrium in \cite{Singh_2007} with no host-feeding, by a factor of 1.23.   

\section{Discussion}\label{sec6}
In this paper, we have considered the classic host-parasitoid interaction with a host-feeding dynamic.  For simplicity and to obtain analytical results, our model ignores gut capacity and includes two states, eggless and with egg, where eggs are produced immediately after consumption.  In contrast to previous phenomenological models, we incorporate the semi-discrete framework, which has more relevance to parasitoid populations with one year life cycles and allows us to track the change in hosts during the vulnerable period.  The preceding analyses show that the effects of host-feeding alone cannot stabilize the classic Nicholson-Bailey model.  However, including density-dependence and a quadratic functional response coupled with the host-feeding dynamic provides a more stabilizing effect as compared to Singh et.~al.'s results \cite{Singh_2007}.  For example, when $R=2$ a stable host-parasitoid equilibrium exists in a wider range of density-dependent mortality rates by a factor of $1.41$ so that for higher values of density-dependent mortality, stability still occurs.  Indeed, in this case, the host-feeding mortality rate is $c_1 = 0.4549$ and without host-feeding, the mortality rate is $c_1 = 0.3466$.  This means host-feeding relaxes the effect of density-dependent mortality and allows higher rates of mortality to exist in the interaction.  Furthermore, the mean host density is increased by a factor of 1.54 in the host-parasitoid stability region.  Therefore, host-feeding makes the parasitoid less efficient in reducing the number of hosts in every generation.  This may seem contradictory since host-feeding can be viewed as an extra process that eliminates the hosts; however, the parasitoid feeds on what could have been a viable host for reproduction in the next generation.  Therefore, host-feeding may have a short term biological control effect but ultimately it reduces the parasitoid population in the long term.  In the case of a quadratic functional response in the attack rates, Singh et.~al.~show that the sole equilibrium point is neutrally stable \cite{Singh_2007}.  In our model, host-feeding stabilizes this equilibrium point and also shows an increase in the mean host density by a factor of 1.23.  Hence, host-feeding brings stability to both systems by decreasing the efficiency of the parasitoid to reproduce from year to year.  

Investigating other key assumptions in our model may lead to further insight into the host-feeding dynamic.  For instance, we assume parasitoids emerge into the vulnerable period each year with an egg.  As Jervis et.~al.~explain in \cite{Jervis_2008}, the adult parasitoid female emerges with eggs, and later feeding on hosts to gain further eggs.  However, do all females enter the vulnerable period with eggs?  Dieckhoff et.~al.~and others conclude that many psychological and behavioral factors including resorption and nutrition may contribute to the specific egg load at any given time \cite{Dieckhoff_2010}.  This translates to the initial condition of our continuous model.  We investigate this by considering a proportion of the starting population that is initially eggless, i.e., $P_0(0) = \lambda P_t$ and $P_1(0) = (1-\lambda) P_t$.  We found that as $\lambda \to 1$ (poor conditions for full egg load at adult emergence), both models are less stable.  Indeed, for $\lambda = 1/2$, it can be shown that Equations \eqref{Mort_1} -- \eqref{Mort_4} yield an identical update to Singh et.~al.'s less stable, host-mortality model, which is the basis for our comparison in Figure \ref{lambda005_plot}.  

Another investigation we considered is a density-dependent parasitoid mortality.  Similar to the model presented in Section \ref{sec4}, we consider a parasitoid death rate that is dependent on the total parasitoid population.  In this sense, our reaction scheme is 
\begin{align}
&P_1 + L \xrightarrow{ c } I + P_0\\
& P_0 + L \xrightarrow{ c } P_{1} \\
& P_0 \xrightarrow{ c_2(P_0 + P_1) }\text{Death} \\ 
& P_1 \xrightarrow{ c_2(P_0 + P_1) }\text{Death},\end{align}
where $c$ represents the constant attack rates by parasitoids as before and $c_2$ is a constant.  This model is analogous to Singh et.~al.'s density-dependent, parasitoid mortality model without host-feeding \cite{Singh_2007}.  As shown by Singh et.~al., the non-host-feeding system is stable if $c_2>c$.  In the host-feeding case, an analytical solution for $I(\tau,t)$ cannot be obtained; however, numerical investigation suggests that the system is stable for $c_2$ values that are slightly less than $c$.  Indeed, a trajectory with $R = 2$, $c = 0.01$, and $c_2 = 0.0098$ yields asymptotically stable results whereas the model without host-feeding experiences diverging oscillations.  Although the stability region is expanded modestly, we can ultimately conclude that host-feeding has a stabilizing effect in the density-dependent parasitoid mortality case.  

\begin{figure}[b!]
\centering
\includegraphics[width = \textwidth]{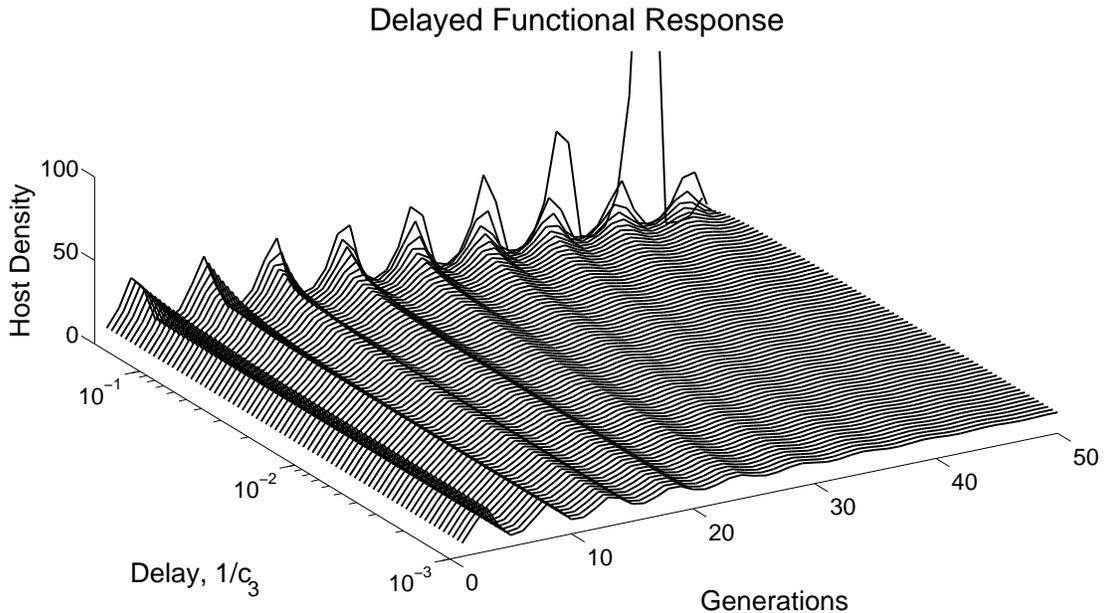}
\captionof{figure}{\textbf{Incorporating a delay due to egg production has a destabilizing effect on the quadratic functional response model.}  A series of host population density ($z$-axis) trajectories are plotted against time ($x$-axis) for each value of egg maturation delay ($y$-axis).  As the delay time, $1/c_3$, approaches $0.18$ or approximately 16 days, the system becomes unstable.  However, the system is always stable for smaller (on the order of hours) delay time.  Simulations are run using continuous system given by Equations \eqref{Func_delay_1} - \eqref{Func_delay_5} and discrete yearly update given by Equations \eqref{Update_F} and \eqref{Update_G} with $T = 1$ and $k = 1$.  Parameters:  $R = 2$, $c = 0.01$, $H(0) = 8$, and $P(0) = 5$.   }
\label{Delay_Functional}
\end{figure}

The study by Shea et.~al.~\cite{Shea_1996} focused on the effects of egg limitation \cite{Reeve_1985} in a host-feeding interaction.  They conclude that stability is effected by the length of latent period, i.e., the time it takes for eggs to mature as the gut is emptied.  Indeed, the longer the latent period, the longer it takes for the system to stabilize.  Among other conclusions, they also observe that stability is not effected by the number of eggs that can be stored.  In this sense, our model considers an instantaneous egg maturation process and parasitoids carry a single egg.  Because we only consider a single egg, we are able to study the system analytically and gather the results shown above.  However, to discuss the case of egg maturation, we implement a delay in egg production into the semi-discrete framework with the following reactions  
\begin{align}
&P_1 + L \xrightarrow{cL} I + P_0\\
& P_0 + L \xrightarrow{cL} P_{1/2} \\
& P_{1/2} \xrightarrow{c_3} P_1.\end{align}
As compared to the original model in Section \ref{sec2}, we add a population, $P_{1/2}$, that is analogous to Shea et.~al.'s~$P_{01}$ population, which represents a parasitoid with zero eggs and a full gut.  The constant rate $c_3$ measures the transition from eggless to with egg as the parasitoid gains enough energy to effectively mature its egg.  Using these reactions, our equations become 
\begin{align}
\de{L}{\tau} & = -c(P_0 + P_1)L^2 \label{Func_delay_1}\\
\de{I}{\tau} & = cP_1L^2 \label{Func_delay_2}\\
\de{P_0}{\tau} & = c(P_1-P_0)L^2 \label{Func_delay_3}\\
\de{P_{1/2}}{\tau} & = cP_1L^2 - c_3P_{1/2}\label{Func_delay_4}\\
\de{P_1}{\tau} & = c_3P_{1/2} - cP_1L^2.\label{Func_delay_5}\end{align}
Analytical solutions to this model are unattainable, but we confirm the destabilizing effect of the latent period numerically.  In Figure \ref{Delay_Functional}, we see a series of host densities using the quadratic functional response model plotted against time with a varying delay parameter, $c_3$.  As the rate of maturation gets slower, the delay increases.  For a critical delay the stable host-feeding, functional response model becomes unstable as the density of hosts begins to experience diverging oscillations.

It is worthwhile to note that the characteristic time scale, $T$, in which the vulnerable period takes place is approximately 90 days.  The instability is caused by a delay that is approximately $0.18$, which suggests that the egg maturation delay is approximately 16 days, which is far too long for an egg to mature.  As illustrated in Figure \ref{Delay_Functional}, a delay value closer to $10^{-2}$ or $10^{-3}$, which means it takes on the order of hours for an egg to mature, yields stable results.  Therefore, in this more realistic case, the system is always stable.

Finally, we note that many other important consumer-resource dynamics can be implemented into the semi-discrete framework.  A necessary future step would be to investigate the effects of a larger egg load and/or to implement a probability that a parasitoid will host feed.  In this sense, an optimal stability criterion could exist.  Furthermore, we seek to investigate susceptibility of host risk to parasitoid attack, as in \cite{Singh_2009}.  Susceptibility of risk cannot stabilize the model with host-feeding alone, but could provide interesting results with host-dependent mortality or a quadratic functional response.  The generality of the semi-discrete framework allows us to implement these changes with ease.



\section{Acknowledgements} 
The authors would like to thank Bill Murdoch and Roger Nisbet for their helpful discussions.  Brooks Emerick would like to thank his advisor, Dr.~Gilberto Schleiniger, for his continued support, and Zhenyu He and Longfei Li for helpful discussions.  Also, A. Singh would like to acknowledge the support from the National Science Foundation Grant DMS-1312926, University of Delaware Research Foundation (UDRF) and Oak Ridge Associated Universities (ORAU).



\bibliography{Parasitoid_Bibliography.bib}
\bibliographystyle{unsrt}

 \appendix
\section{Analysis of Density Independent Mortality}\label{appA}
We consider the explicit solution to Equations \eqref{No_Mort_1} -- \eqref{No_Mort_4}.  Adding \eqref{No_Mort_3} and \eqref{No_Mort_4} gives  
\begin{equation} \de{P_0}{\tau} + \de{P_1}{\tau} = 0 \qquad \Rightarrow \qquad P_0 + P_1 = P_t.\end{equation}
\noindent Substituting $P_1 = P_t - P_0$ into \eqref{No_Mort_1} yields 
\begin{equation} \de{L}{\tau} = -c P_t L \qquad \Rightarrow \qquad L(\tau,t) = RH_t \text{exp} \left(-cP_t \tau\right).\end{equation}
We can now solve for $P_0$ by substituting our expression for $L$ and $P_1$ into equation \eqref{No_Mort_3} to obtain 
\begin{equation} \de{P_0}{\tau} + 2cRH_t\exp(-cP_t\tau)\, P_0= cRH_tP_t  \text{exp}\left(-c P_t \tau\right).\end{equation}
Using the integrating factor to solve this equation, we get an expression for both $P_0$ and $P_1$.  We have 
\begin{equation} P_0(\tau,t) = \frac{P_t}{2}- \frac{P_t}{2} \,\text{exp}\left\{ \frac{RH_t}{P_t}\big[ 1 - \text{exp}\left( - cP_t \tau\right)\big] \right\}\end{equation}
\begin{equation} P_1(\tau,t) = \frac{P_t}{2} + \frac{P_t}{2} \,\text{exp}\left\{ \frac{RH_t}{P_t}\big[ 1 - \text{exp}\left( - cP_t \tau\right)\big] \right\}.\end{equation}
Finally, substituting our expression for $P_1$ into equation \eqref{No_Mort_2} gives 
\begin{equation} \de{I}{\tau} = \frac{cRH_tP_t}{2}\text{exp}\left( -c R_t \tau\right) + \frac{cRH_tP_t}{2}\, \text{exp}\left(-cP_t\tau\right) \text{exp}\left\{ \frac{R H_t}{P_t} \big[ 1 - \text{exp}\left(-cP_t \tau\right)\big] \right\}.\end{equation}
Solving this and applying the initial condition gives the following expression for $I(\tau,t)$ 
\begin{equation} I(\tau,t) = \frac{RH_t}{2} \big[ 1 - \text{exp} \left(-cP_t \tau\right) \big] + \frac{P_t}{2}\, \left\{ \text{exp}\left\{ \frac{RH_t}{P_t} \big[ 1 - \text{exp}\left(-cP_t\tau\right) \big] \right\} - 1 \right\}.   \end{equation}
Using the solutions for $L$ and $I$, we find the update as in Equations \eqref{No_Mort_Update_H} and \eqref{No_Mort_Update_P} using the definition in Equations \eqref{Update_F} and \eqref{Update_G}.  We can analyze the resulting discrete-time model by implementing the procedure outlined in \cite{Elaydi_1996}.  Using a general discrete model such as 
\begin{align}
H_{t+1} & = F(H_t, P_t) \\
P_{t+1} & = G(H_t, P_t),\end{align}
we can perform a linear stability analysis about the fixed point $(H^*, P^*)$, where $H^*$ and $P^*$ satisfy the following system of equations 
\begin{align}
H^*& = F(H^*, P^*) \\
P^* & = G(H^*, P^*), \end{align}
by determining if the spectral radius of the Jacobian matrix evaluated at the fixed point is less than one.  That is, if the magnitude of the eigenvalues of the following Jacobian matrix,
\begin{equation} J := J(H^*, P^*) = \left. \begin{bmatrix} \left. \pd{F}{H} \right|_{(H^*, P^*)} & \left. \pd{F}{P}\right|_{(H^*, P^*)} \\ \left.\pd{G}{H}\right|_{(H^*, P^*)} & \left.\pd{G}{P}\right|_{(H^*, P^*)} \end{bmatrix}\right. ,\end{equation}
are within the unit circle, then the fixed point $(H^*, P^*)$ is asymptotically stable.  The eigenvalues fall within the unit circle if the following three Jury conditions hold, 
\begin{align}
1 - \text{Tr}(J) + \text{Det}(J) & >0\\
1 + \text{Tr}(J) + \text{Det}(J) & >0\\
1 - \text{Det}(J) & >0.\end{align}
In the case of density independent mortality, our update functions are given by Equations \eqref{No_Mort_Update_H} and \eqref{No_Mort_Update_P}, 
\begin{align}
F(H_t, P_t) &= RH_t e^{-cP_t }\\
G(H_t, P_t) &= \frac{RH_t}{2} \big( 1 - e^{-cP_t } \big) +  \frac{P_t}{2}\, \left\{ \text{exp}\left[ \frac{RH_t}{P_t} \big( 1 - e^{-cP_t} \big) \right] - 1 \right\}, \end{align}
with fixed point 
\begin{equation} H^* = \frac{\beta \ln R}{c(R-1)}, \qquad P^* = \frac{\ln R}{c}, \qquad \beta = 0.7921.\end{equation}
Using these expressions, we can evaluate the trace and determinant of the Jacobian matrix as 
\begin{align}
\text{Tr}(J) & =  \frac{\beta(1 + e^{\beta}) }{2} \left[ \frac{e^\beta(1-\beta) + 1}{\beta(1+e^\beta )} + \frac{\ln R}{ R-1} \right]\\
\text{Det}(J) & = \frac{\beta(1 + e^{\beta}) }{2} \left[ \frac{e^\beta(1-\beta) - 1}{\beta(1+e^\beta )} + \frac{\ln R}{ R-1} \right]. \end{align}
We can see that $1 - \text{Tr}(J) + \text{Det}(J) = 0$ for all $R$.  Hence, the fixed point is not asymptotically stable.  

\section{Analysis of density-dependent Mortality Model}\label{appB}
We consider the explicit solution to Equations \eqref{Mort_1} -- \eqref{Mort_4}.  Adding \eqref{Mort_3} and \eqref{Mort_4} gives  $P_0 + P_1 = P_t$, which upon substitution of $P_1 = P_t - P_0$ into Equation \eqref{Mort_1} yields 
\begin{equation} \de{L}{\tau} = \left(-c P_t + c_1L\right) L \qquad \Rightarrow \qquad L(\tau,t) = \frac{RH_t\exp(-c P_t \tau)}{1 + c_1 RH_t \frac{1-\exp(-c P_t \tau)}{c P_t}}.  \end{equation}
The previous differential equation can be solved using partial fraction decomposition.  We can now solve for $P_0$ by substituting our expression for $L$ and $P_1$ into equation \eqref{Mort_3} to obtain 
\begin{equation} \de{P_0}{\tau} + \frac{2cRH_t  \text{exp}\left(-c P_t \tau\right)}{1 + c_1 RH_t \frac{1-\exp(-c P_t \tau)}{c P_t}}\, P_0 = \frac{cRH_tP_t  \text{exp}\left(-c P_t \tau\right)}{1 + c_1 RH_t \frac{1-\exp(-c P_t \tau)}{c P_t}}.  \end{equation}
Using the integrating factor to solve this equation, we get an expression for both $P_0$ and $P_1$.  Solving this explicitly for $P_0$ and using the initial condition from the previous section yields our solution for both $P_0$ and $P_1$ as 
\begin{equation} P_0(\tau,t) = \frac{P_t}{2} -  \frac{1}{ 2 \left[ 1 + c_1 RH_t \frac{1-\exp(-c P_t \tau)}{c P_t}\right]^{\frac{2c}{c_1}}}  \end{equation}
\begin{equation} P_1(\tau,t) = \frac{P_t}{2} +  \frac{1}{ 2 \left[ 1 + c_1 RH_t \frac{1-\exp(-c P_t \tau)}{c P_t}\right]^{\frac{2c}{c_1}}}.  \end{equation}
Finally, substituting our expression for $P_1$ into equation \eqref{Mort_2} gives 
\begin{equation} \de{I}{\tau} = \frac{cP_t\exp(-c P_t \tau)}{2\left[1 + c_1 RH_t \frac{1-\exp(-c P_t \tau)}{c P_t}\right]} + \frac{c\exp(-c P_t \tau)}{2\left[1 + c_1 RH_t \frac{1-\exp(-c P_t \tau)}{c P_t}\right]^{\frac{2c}{c_1}+1}}.\end{equation}
Solving this and applying the initial condition gives the following expression for $I(\tau,t)$ 
\begin{equation} I(\tau,t) =\frac{cP_t}{2c_1} \ln \left[1 + c_1 RH_t \frac{1-\exp(-c P_t \tau)}{c P_t} \right] - \frac{P_t}{4}\left\{ \left[ 1 + c_1 RH_t \frac{1-\exp(-c P_t \tau)}{c P_t}\right]^{-\frac{2c}{c_1}}-1\right\}.\end{equation}
Using the solutions for $L$ and $I$, we find the update as in Equations \eqref{Mort_Update_H} and \eqref{Mort_Update_P} using the definition in Equations \eqref{Update_F} and \eqref{Update_G}.  In the density-dependent host mortality case, our discrete update functions are,
\begin{align}
F(H_t, P_t) & = \frac{RH_t\exp(-c P_t)}{1 + c_1 RH_t \frac{1-\exp(-c P_t )}{c P_t}}\\
G(H_t, P_t) & = \frac{cP_t}{2c_1} \ln \left[1 + c_1 RH_t \frac{1-\exp(-c P_t )}{c P_t} \right] - \frac{P_t}{4}\left\{ \left[ 1 + c_1 RH_t \frac{1-\exp(-c P_t )}{c P_t}\right]^{-\frac{2c}{c_1}}-1\right\}. \end{align}
This system has two nontrivial fixed points.  As noted in the main text, the first no-parasitoid equilibrium point is given by 
\begin{equation}
H^*_{NC} =  \frac{R-1}{c_1 R}, \qquad P^* = 0.\end{equation}
Denoting the Jacobian matrix for the first fixed point by $J_{NC}$, we find the trace and determinant to be
\begin{align}
\text{Tr}(J_{NC}) & = \frac{1}{R} - \frac{1}{4}\left[ R^{-\frac{2}{z}} - 1  - \frac{2\ln R}{z} \right]\\
\text{Det}(J_{NC}) & = - \frac{1}{4R}\left[ R^{-\frac{2}{z}} - 1  - \frac{2\ln R}{z} \right], \end{align}
where $z = c_1/c$.  Using these expressions, Jury condition 2 gives the most strict condition so that $z>\ln R/\gamma$ for stability.  The second equilibrium point is given by 
\begin{equation}
H^* = \left(\frac{\exp\left( \frac{\gamma c_1}{c} \right) - 1 }{ 1 -\frac{ \exp\left( \frac{\gamma c_1}{c} \right)}{R} }\right) \frac{ c P^* }{ c_1 }, \qquad P^* = \frac{\ln(R) - \frac{\gamma c_1}{c} }{c} , \qquad \gamma = 1.5238. \end{equation}
Define $J$ as the Jacobian matrix evaluated at the second equilibrium point above, then the trace and determinant are given by 
\begin{align}
\text{Tr}(J) & = \frac{\left\{ e^{\gamma z} \left[ \ln R - (\gamma z - 1) \right] - R\right\}(1 - e^{-\gamma z} ) ( 1 + e^{-2\gamma})}{2z (R - e^{\gamma z}) } + \frac{1}{2} \left[ \gamma + \frac{1}{2} \left( 1 - e^{-2\gamma}\right)\right] + e^{-\gamma z} \\
\text{Det}(J) & = \frac{\left\{ R\left[ \ln R - \left(\gamma z + 1\right) \right] \left( 1 - e^{-\gamma z}\right) +  e^{\gamma z}-1 \right\} \left( 1 + e^{-2\gamma}\right) }{2 z \left(R- e^{\gamma z} \right) } + \frac{1}{2} \left[ \gamma + \frac{1}{2} \left( 1- e^{-2\gamma} \right) \right] e^{-\gamma z}. \end{align}
Here, Jury condition 3 is the most strict, and we must have $z^*<z<\ln R/\gamma$, where $z^*$ solves $1-\text{Det}(J)=0$.  Figure \ref{lambda005_plot} shows a plot for both stability regions.

\section{Analysis of Quadratic Functional Response Model}\label{appC}
We consider the explicit solution to Equations \eqref{Func_1} -- \eqref{Func_4}.  Adding \eqref{Func_1} and \eqref{Func_4} gives  $P_0 + P_1 = P_t$.  Substituting $P_1$ into Equation \eqref{Func_1} yields
\begin{equation} \de{L}{\tau} = -c P_t L^2 \qquad \Rightarrow \qquad L(\tau,t) = \frac{RH_t}{1 + cRH_tP_t\tau}.\end{equation}
We can now solve for $P_0$ by substituting our expression for $L$ and $P_1$ into equation \eqref{Func_1} to obtain 
\begin{equation} \de{P_0}{\tau} + \frac{2cR^2H_t^2}{(1+cRH_tP_t \tau)^2} P_0= \frac{ cR^2H_t^2P_t  }{(1 + cRH_tP_t \tau)^2} .\end{equation}
Using the integrating factor to solve this equation, we get an expression for both $P_0$ and $P_1$.  We have 
\begin{equation} P_0(\tau,t) = \frac{P_t}{2}- \frac{P_t}{2} \,\text{exp}\left[ \frac{2RH_t}{P_t}\left( \frac{1}{1 + cRH_t P_t \tau} - 1\right) \right]\end{equation}
\begin{equation} P_1(\tau,t) = \frac{P_t}{2}+ \frac{P_t}{2} \,\text{exp}\left[ \frac{2RH_t}{P_t}\left( \frac{1}{1 + cRH_t P_t \tau} - 1\right) \right].\end{equation}
Substituting our expressions for $P_1$ and $L$ into Equation \eqref{Func_2}, we obtain the following ODE for $I$, 
\begin{equation} \de{I}{\tau}  = \frac{cR^2H_t^2P_t}{2(1 + cRH_tP_t\tau)^2} +  \frac{cR^2H_t^2P_t\exp\left(-\frac{2RH_t}{P_t}\right)\exp\left[ \frac{2RH_t}{P_t(1 + cRH_tP_t \tau)}\right]}{2(1 + cRH_tP_t\tau)^2} .\end{equation}
We can solve this equation directly to obtain the following expression for $I$, 
\begin{equation} I(\tau, t) = \frac{RH_t}{2} \left(1 - \frac{1}{1 + cRH_tP_t \tau} \right) + \frac{P_t}{4} \left\{ 1 - \exp\left[ \frac{2RH_t}{P_t} \left( \frac{1}{1 + cRH_tP_t \tau} - 1\right) \right] \right\}. \end{equation}
Using the solutions for $L$ and $I$, we find the update as in Equations \eqref{Func_Update_H} and \eqref{Func_Update_P} using the definition in Equations \eqref{Update_F} and \eqref{Update_G}.  In the quadratic functional response case, our discrete update functions are, 
\begin{align}
F(H_t, P_t) & =\frac{R H_t}{1 + cRH_tP_t}\\
G(H_t, P_t) & =\frac{RH_t}{2} \left(1 - \frac{1}{1 + cRH_tP_t} \right) + \frac{P_t}{4} \left\{ 1 - \exp\left[ \frac{2RH_t}{P_t} \left( \frac{1}{1 + cRH_tP_t } - 1\right) \right] \right\},\end{align}
which gives rise to the following fixed point
\begin{equation} H^* = \sqrt{\frac{\gamma}{cR}}, \qquad P^* = \frac{R - 1}{\sqrt{\gamma cR}}, \qquad \gamma = 1.5238.\end{equation}
The trace and determinant of the Jacobian matrix is given by 
\begin{align}
\text{Tr}(J) & = \frac{1}{4} - \frac{e^{-2\gamma}}{2} \left( \gamma + \frac{1}{2}\right)  + \frac{1}{R} \left[ 1 + \frac{\gamma}{2} \left( 1 + e^{-2\gamma} \right) \right]\\
\text{Det}(J) & =\frac{\gamma}{2} \left( 1 + e^{-2\gamma} \right) + \frac{1}{4R} \left[ 1 - (1 + 2\gamma) e^{-2\gamma} \right]. \end{align}
All three Jury conditions hold for $R>1$.

\end{document}